\documentclass[%
 reprint,
 amsmath,amssymb,
 aps,
aps,prb,reprint,floatfix,superscriptaddress]{revtex4-2}

\usepackage{graphicx}
\usepackage{float}
\usepackage{dcolumn}
\usepackage{bm}

\usepackage{lipsum}
\begin{document}

\preprint{APS/123-QED}

\title{Odd-parity Magnetism from the Generalized Bloch Theorem}
\author{Mikkel Christian Larsen}
%
\author{Thomas Olsen}
\thanks{Corresponding author. Email: \href{mailto:tolsen@fysik.dtu.dk}{tolsen@fysik.dtu.dk}}
\affiliation{%
 Department of Physics, Technical University of Denmark, 2800 Kgs. Lyngby Denmark
}%
\date{\today}

\begin{abstract}
In the non-relativistic limit, helimagnetic order is always associated with odd-parity magnetism. That is, for single-particle states the expectation value of the electronic spin is odd in crystal momentum, which implies direct control of the spin by means of electric fields. However, the theoretical description of helimagnets is hindered by the fact that the spiral pitch may require large super cells or even be incommensurate with the lattice. In the this letter we show that such issues may be remedied by use of the Generalized Bloch theorem. It allows one to describe (by models or first principles) the system in terms of the primitive unit cell, from which all relevant properties can be obtained by downfolding in reciprocal space. We exemplify the procedure using MnI$_2$ and NiI$_2$, which are known type II multiferroics having spiral order and the helimagnetic metal MnTe$_2$. We analyze how the magnitude of spin splitting depends on orbital composition of bands, and we show that spin splitting is maximized for states having large odd-orbital ($p$-type) character. It is straightforward to generalize the framework to handle response functions for helimagnets using only the primitive unit cell and the present downfolding procedure thus strongly facilitate theoretical progress in the field.
\end{abstract}

\maketitle

Lifting the spin degeneracy of electronic bands is crucial for manipulating spin currents in solid state devices, and the resulting energy splitting plays a prominent role in spintronics technology \cite{zutic_spintronics_2004, jungwirth_multiple_2018, yakout_spintronics_2020}. Moreover, if the electronic spin is entangled with its momentum, it becomes possible to control the spin by means of an external electric field, which facilitates spin-to-charge conversion in spintronics devices. The Rashba splitting is one example of such an effect \cite{manchon_new_2015}, which is typically induced by introducing a surface or interface that breaks inversion symmetry. However, this effect is derived from spin-orbit coupling and the magnitude of spin splitting is thus limited by the strength of intrinsic spin-orbit effects in the constituent elements. This has initiated a vast interest in altermagnetism, which denotes collinear spin-compensated magnetic order exhibiting {\it non-relativistic} spin splitting \cite{yuan_giant_2020,smejkal_beyond_2022, smejkal_emerging_2022}. Such splitting may become much larger than the corresponding one driven by spin-orbit effects, and altermagnetism has thus been proposed as an ideal platform for designing efficient spintronics devices \cite{smejkal_emerging_2022}. Altermagnets have even parity in the sense that an electron at momentum $\mathbf{k}$ will have the same spin as one at $-\mathbf{k}$, and they can be classified as being $d$-wave, $g$-wave, etc. depending on how the spin transforms upon rotation of the momentum. While such symmetric spin-momentum locking may be useful for a wide range of purposes, materials with anti-symmetric spin-momentum locking open a distinct new set of possible functionality - for example the non-relativistic Edelstein and torque effects \cite{gonzalez-hernandez_non-relativistic_2024}. It is not possible to realize antisymmetric (odd) spin-momentum locking in collinear magnets, but certain non-collinear magnets may exhibit exactly such intrinsic odd-parity magnetism \cite{hellenes_p-wave_2024, chakraborty_highly_2025, yu_odd-parity_2025, yamada_metallic_2025}. In particular, it is straightforward to show that any coplanar (and non-collinear) single-$q$ state, will exhibit odd-parity magnetism. However, theoretical modeling of such states is strongly inhibited by the large super cells required to encompass the spiral wave-vector $\mathbf{Q}$ (which may even be incompatible with the lattice).

In the present work we show that the Generalized Bloch theorem (GBT) \cite{sandratskii_symmetry_1991, heide_describing_2009, sodequist_magnetic_2024} is ideally suited to handle single-$q$ states exhibiting odd-parity magnetism. Specifically, we show that the bands can be calculated in the primitive (crystallographic) unit cells, and we further derive the downfolding procedure required to obtain the bands in the super cell defined by the magnetic ordering vector $\mathbf{q}$. The procedure is exemplified by performing first principles calculations on the two-dimensional materials MnI$_2$ and NiI$_2$, which have recently gained attention due to their multiferroic properties \cite{song_evidence_2022, song_electrical_2025}, and the odd-parity metal MnTe$_2$. 
We also show that $p$-character of electronic bands is crucial for having finite spin polarization.
\begin{figure*}
\centering
\includegraphics[width = \textwidth, trim=0.2cm 4cm 0cm 3cm, clip]{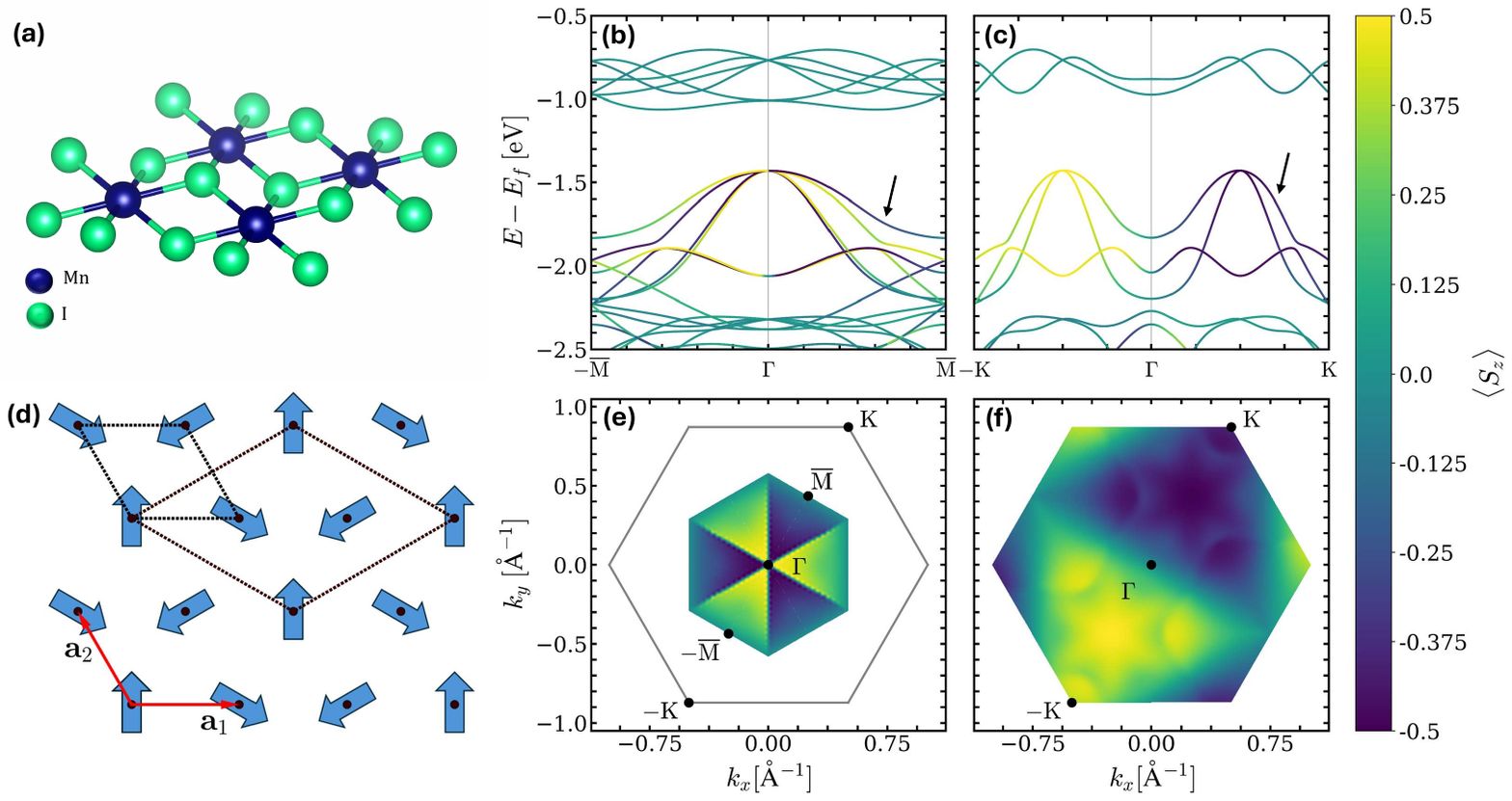}
\caption{\label{fig:MnI2_results} Application of Blochs theorem and associated downfolding exemplified by hexagonal $\mathrm{MnI_2}$. (\textbf{a}) crystal structure, (\textbf{b}) band structure in super cell with energy relative to the Fermi-energy, (\textbf{c}) band structure in primitive cell with GBT with energy relative to the Fermi-energy, (\textbf{d}) magnetic moments in super cell and primitive cell, (\textbf{e}) spin expectation value within the magnetic Brillouin zone, and (\textbf{f}) spin expectation value within the crystallographic Brillouin zone. The bands chosen in (\textbf{e}) and (\textbf{f}) are indicated by arrows in (\textbf{b}) and (\textbf{c}).\label{MnI2_bs} }
\end{figure*}
We consider the general case of a coplanar single-$q$ state, which can be characterized by the magnetization density satisfying $\mathbf{m}(\mathbf{r}+\mathbf{r}_i)=R_\mathbf{\hat n}(\mathbf{q}\cdot\mathbf{r}_i)\mathbf{m}(\mathbf{r})$, where $\mathbf{r}_i$ is a lattice vector of the primitive (crystallographic) cell, and $R_\mathbf{\hat n}(\mathbf{q}\cdot\mathbf{r}_i)$ rotates the magnetization by an angle $\varphi=\mathbf{q}\cdot\mathbf{r}_i$ around the normal vector $\mathbf{\hat n}$ of the spiral plane. It is straightforward to demonstrate that such a configuration will exhibit odd-parity magnetism. This was shown in Ref. \cite{gonzalez-hernandez_non-relativistic_2024} and below we provide a brief and slightly alternative argument. In a mean field treatment, the Hamiltonian $H$ (without spin-orbit coupling) will exhibit the combined symmetry of lattice translation and spinor rotation $U_\mathbf{q}(\mathbf{r}_i)$. In the magnetic unit cell (super cell) the Bloch eigenstates are characterized by a crystal momentum $\mathbf{K}$, and for a spin operator $\mathbf{S}_\perp$ orthogonal to $\mathbf{\hat n}$ these will satisfy $\langle\psi_\mathbf{K}|\mathbf{ S}_\perp|\psi_\mathbf{K}\rangle=\langle\psi_\mathbf{K}|U^\dag_\mathbf{q}(\mathbf{r}_i)\mathbf{S}_\perp U_\mathbf{q}(\mathbf{r}_i)|\psi_\mathbf{K}\rangle=\langle\psi_\mathbf{K}|R_\mathbf{\hat n}(\mathbf{q}\cdot\mathbf{r}_i)\mathbf{S}_\perp|\psi_\mathbf{K}\rangle$, which implies that the expectation value must vanish (for a non-degenerate state). On the other hand, for a spin operator orthogonal to the plane one obtains $\langle\psi_\mathbf{K}|\mathbf{\hat n}\cdot\mathbf{S}|\psi_\mathbf{K}\rangle=-\langle\psi_\mathbf{-K}|\mathbf{\hat n}\cdot\mathbf{S}|\psi_\mathbf{-K}\rangle$, and in general such states will have $\varepsilon_\mathbf{-K}=\varepsilon_\mathbf{K}$. This is most easily seen if the out-plane direction is chosen as the $y$-direction, in which case the magnetization density is in the $xz$-plane, and the Hamiltonian is real. Since for complex conjugation $K$ we have that $K\psi_\mathbf{K}(\mathbf{r})\propto\psi_\mathbf{-K}(\mathbf{r})$, and the states at $\mathbf{K}$ and $-\mathbf{K}$ must be degenerate. In addition, $\langle\psi_\mathbf{K}|S_y|\psi_\mathbf{K}\rangle=-\langle\psi_\mathbf{K}|KS_yK|\psi_\mathbf{K}\rangle=-\langle\psi_\mathbf{-K}|S_y|\psi_\mathbf{-K}\rangle$, which implies that the out-of-plane spin direction is odd with respect to the crystal momentum. We note that if $R_\mathbf{\hat n}(\mathbf{q}\cdot\mathbf{r}_i)$ generates a $\pi$ rotation, it will enforce two-fold spin degeneracy of any state at a particular $K$-point, and non-relativistic spin splitting will be absent.

The combined translational and spin rotational symmetry may also be utilized to write eigenspinors as generalized Bloch states:
\begin{align}
\label{Bloch_generalized}
\psi_{\mathbf{q}, \mathbf{k}}(\mathbf{r})
= \mathrm{e}^{i\mathbf{k}\cdot\mathbf{r}}\,
U_{\mathbf{q}}^\dagger(\mathbf{r})
u_{\mathbf{q},\mathbf{k}}(\mathbf{r})
\end{align}
where $u_{\mathbf{q},\mathbf{k}}$ are spinors periodic in the crystallographic unit cell (primitive cell) and $\mathbf{k}$ is the crystal momentum, which takes values in the Brillouin zone corresponding to the primitive cell. The spin rotation matrix is given by
\begin{equation}
    U_{\mathbf{q}}(\mathbf{r}) = \begin{bmatrix}
        \mathrm{e}^{i\mathbf{q}\cdot \mathbf{r}/2} & 0 \\
        0& \mathrm{e}^{-i\mathbf{q}\cdot \mathbf{r}/2}
    \end{bmatrix},
\end{equation}
in a basis of spinors along $\mathbf{\hat n}$.
The eigenvalue problem can then be formulated in the primitive cell as
\begin{equation}
    H_{\mathbf{q}, \mathbf{k}}u_{\mathbf{q}, \mathbf{k}} = \varepsilon_{\mathbf{q}, \mathbf{k}}u_{\mathbf{q}, \mathbf{k}}
\end{equation}
with the generalized Bloch Hamiltonian given by
\begin{equation}
    H_{\mathbf{q}, \mathbf{k}} = \mathrm{e}^{-i\mathbf{k}\cdot \mathbf{r}} U_{\mathbf{q}}(\mathbf{r}) H  U^\dagger_{\mathbf{q}}(\mathbf{r}) \mathrm{e}^{i\mathbf{k}\cdot \mathbf{r}}.
\end{equation}
In contrast, the ordinary Bloch states are written as
\begin{align}
\label{Bloch}
\tilde \psi_{\mathbf{K}}(\mathbf{r})
= \mathrm{e}^{i\mathbf{K}\cdot\mathbf{r}}\,
\tilde u_{\mathbf{K}}(\mathbf{r}),
\end{align}
where $\mathbf{K}$ is in the Brillouin zone corresponding to the
lattice vectors $\mathbf{R}_i$, which generate the magnetic lattice and satisfy $\mathbf{q}\cdot\mathbf{R}_i=2\pi n$, with $n$ being an integer. These states must be simply related to the generalized Bloch states, since both are eigenstates of the combined spin rotation and (fractional) translation. Specifically, requiring that the eigenvalues under translation of a super cell lattice vector are identical yields
\begin{equation}
\label{Bloch-phase}
    (\mathbf{K} - \mathbf{k} + \mathbf{q}/2) \cdot \mathbf{R}_i=   2\pi n,
\end{equation}
where we used that $e^{i\mathbf{q}\cdot\mathbf{R}_i/2}=e^{-i\mathbf{q}\cdot\mathbf{R}_i/2}=\pm1$. Importantly, since $\mathbf{\hat n}\cdot\mathbf{S}$ commutes with $U_{\mathbf{q}}(\mathbf{r})$ one may evaluate the expectation value $\langle\psi_\mathbf{K}|\mathbf{\hat n}\cdot\mathbf{S}|\psi_\mathbf{K}\rangle$ directly from the generalized Bloch state $\psi_\mathbf{k}$ that downfolds to $\mathbf{K}$.

We exemplify the spin splitting and downfolding procedure for helimagnets through calculations of the electronic band structure using Density Functional Theory. The calculations here were performed  with the electronic structure package GPAW \cite{mortensen_gpaw_2024} using the LDA exchange-correlation functional \cite{perdew_accurate_1992} and a plane wave cutoff of 600 eV. As a first example, we consider two-dimensional (2D) MnI$_2$ \cite{, mcguire_crystal_2017}, in which DFT predicts the Mn atoms to comprise a $S=5/2$ triangular lattice with $\mathbf{Q}=(1/3,1/3)$ helimagnetic order \cite{xiang_general_2011,sodequist_type_2023}. This structure is also observed experimentally in magnetic fields exceeding 3 T \cite{kurumaji_magnetic-field_2011}. The crystal structure of MnI$_2$ and the ordered Mn magnetic moments are shown in Fig. \ref{fig:MnI2_results}, where we also indicate the magnetic unit cell, which is three times larger than the crystallographic one. We then calculate the band structure using the primitive cell and the generalized Bloch theorem with $\mathbf{Q}=(1/3,1/3)$ periodic boundary conditions. The result is shown in Fig. \ref{fig:MnI2_results} for a single-band (colored according to spin expectation value) and for all bands along the $-\mathrm{K}\Gamma\mathrm{K}$ line. In this case $\mathbf{Q}$ coincides with the point $\mathrm{K}$, and we clearly observe a nodal line orthogonal to $\mathbf{Q}$ rendering the spin structure "$p$-like". The band structure of the magnetic cell can then be obtained by downfolding according to Eq. \eqref{Bloch-phase}. For example, the band maxima located at $\pm K/2$ obtained with the GBT downfolds to $\Gamma$, since $\mathbf{Q}/2$ is located at $\bar M$ and $\pm K/2$ is at $\pm \bar M$. Similarly, the $\Gamma$-point and $\pm K$ of the primitive BZ downfolds to $\bar M$. The spin clearly exhibits $f$-wave symmetry in the magnetic BZ, and in this particular case the $p$-wave symmetry of the primitive BZ is an artifact of the GBT construct. In fact, the magnetic structure may equivalently be represented by $\mathbf{Q}=(1/3, -2/3)$, which would yield a spin structure with a different nodal line (again orthogonal to  $\mathbf{Q}$), but the downfolded bands and spin are identical to that shown in Fig. \ref{fig:MnI2_results}. We note that $f$-wave symmetry is expected for the $\mathbf{Q}=(1/3, 1/3)$ order because the {\it magnetic} ground state has three-fold symmetry, which is incompatible with $p$-wave order. Generic (non-zero) ordering vectors on the triangular lattice lack such rotational symmetry and therefore typically result in $p$-wave order.
\begin{figure}[t]
\centering
\includegraphics[width=0.9\linewidth]{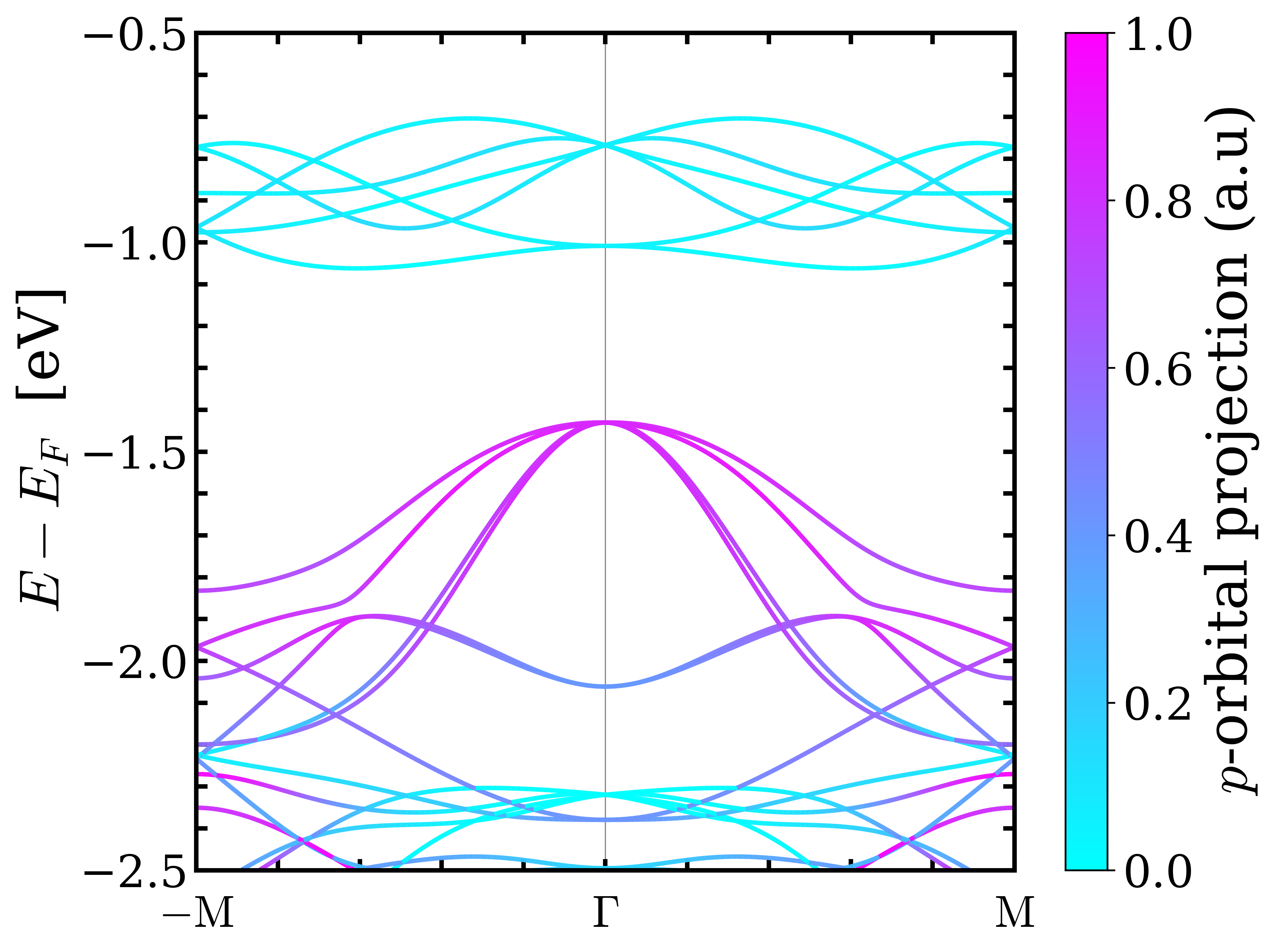}
\caption{\label{MnI2_dos} Projected $p$-orbital character of magnetic ions (Mn).
The $p$-wave character is averaged over the three Mn atoms in
the super cell.}
\end{figure}

While $\langle\psi_\mathbf{K}|S_z|\psi_\mathbf{K}\rangle$ is guaranteed to be odd under inversion of $\mathbf{K}$, the magnitude of the expectation value can be much smaller than the nominal value of $\hbar/2$. In particular, for MnI$_2$, the 6 bands located between -1.1 and -0.7 eV has $\langle S_z\rangle\approx0$, while the bands below has spin expectation values of the order $\hbar/2$. Since any technological application of spin split bands will rely on controlling particular spin states, it is crucial to understand the conditions that lead to states carrying finite spin. In Fig. \ref{MnI2_dos} we show the band structure colored according to the $p$-orbital character. It is clear that the $p$-character (odd) is strongly correlated with the magnitude of spin, which implies that only states composed of odd-parity orbitals show significant spin polarization. In addition, the spin splitting has a significance dependence the ordering vector $\mathbf{Q}$. Although the ground state of MnI$_2$ is at $\mathbf{Q}=(1/3, 1/3)$ we can use the GBT to investigate how the splitting depends on $\mathbf{Q}$. Doing this reveals a maximal splitting at $\mathbf{Q}=(1/3, 1/3)$ . This is rationalized by noting that for $\mathbf{Q}=(0, 0)$ the spin polarization vanished and at $\mathbf{Q}=(0, \pm1/2)$, the splitting vanishes due to collinear bands.


\begin{figure}[t]
\centering
\includegraphics[width=0.99\linewidth]{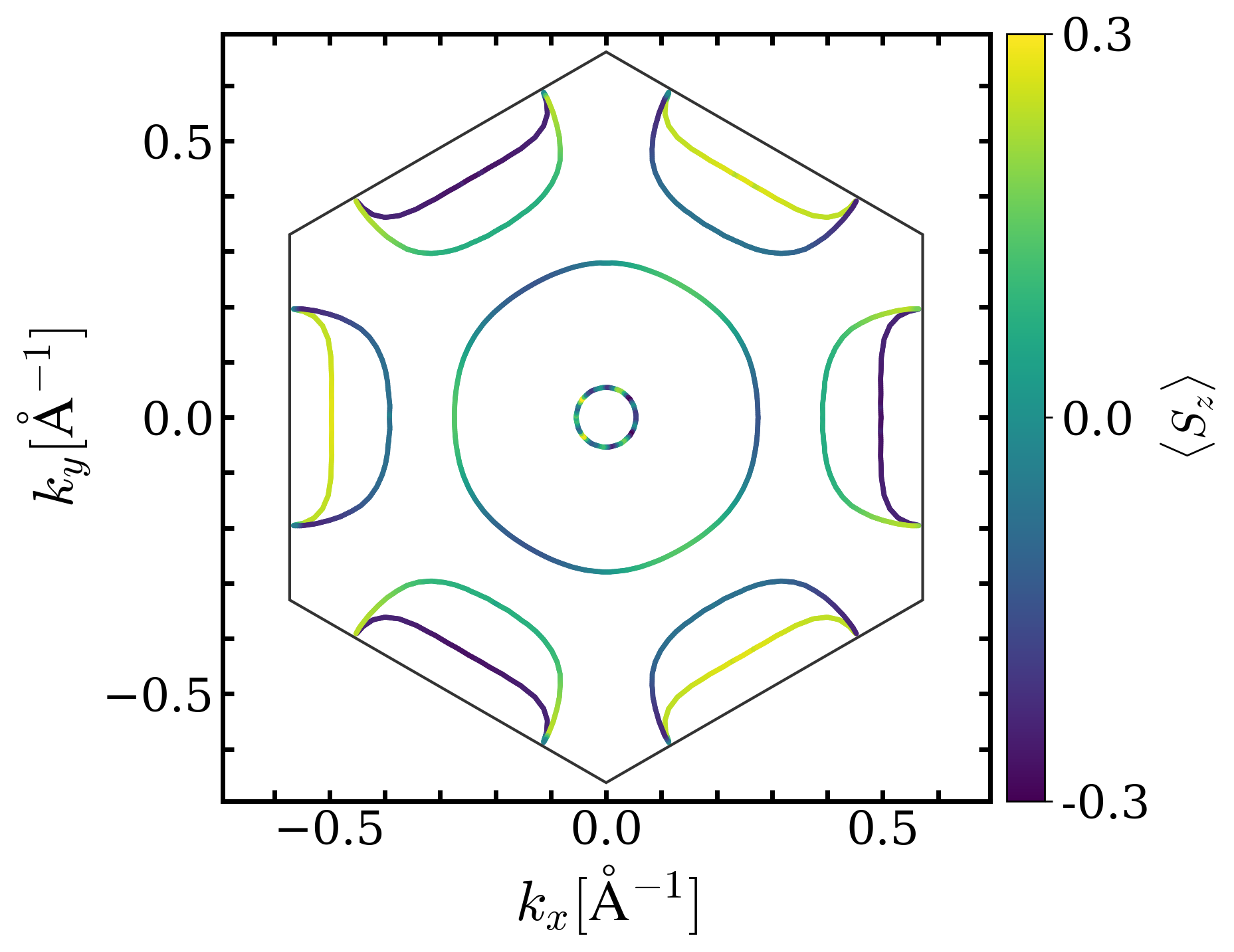}
\caption{\label{fig:Fermi_Surface}$\mathrm{MnTe_2}$ Fermi surface colored according to spin expectation value. The calculation is made for the super cell corresponding to $\mathbf{Q}=(1/3, 1/3)$. }
\end{figure}
Spintronic applications typically rely on conductive spin-polarized electrons and the Fermi surface thus constitutes the main object of interest. As an example of this we consider 2D $\mathrm{MnTe_2}$ in the 2H phase (layer group $p\bar6m2$), which is predicted to be metallic \cite{haastrup_computational_2018} and have $\mathbf{Q}=(1/3, 1/3)$ ground state order \cite{sodequist_magnetic_2024}. 
In Fig.~\ref{fig:Fermi_Surface}, we show the Fermi surface colored according to the spin expectation value for the super cell. It is clear that the Fermi surface exhibits odd-parity, with three--fold symmetry ($f$-wave order) as expected from the space group. 
We also observe the nodal lines to be orthogonal to $\mathbf{Q}$ similar to the GBT calculations for $\mathrm{MnI_2}$ above.

\begin{figure}[t]
\centering
\includegraphics[width=0.99\columnwidth, trim=0.0cm 3cm 0cm 2cm, clip]{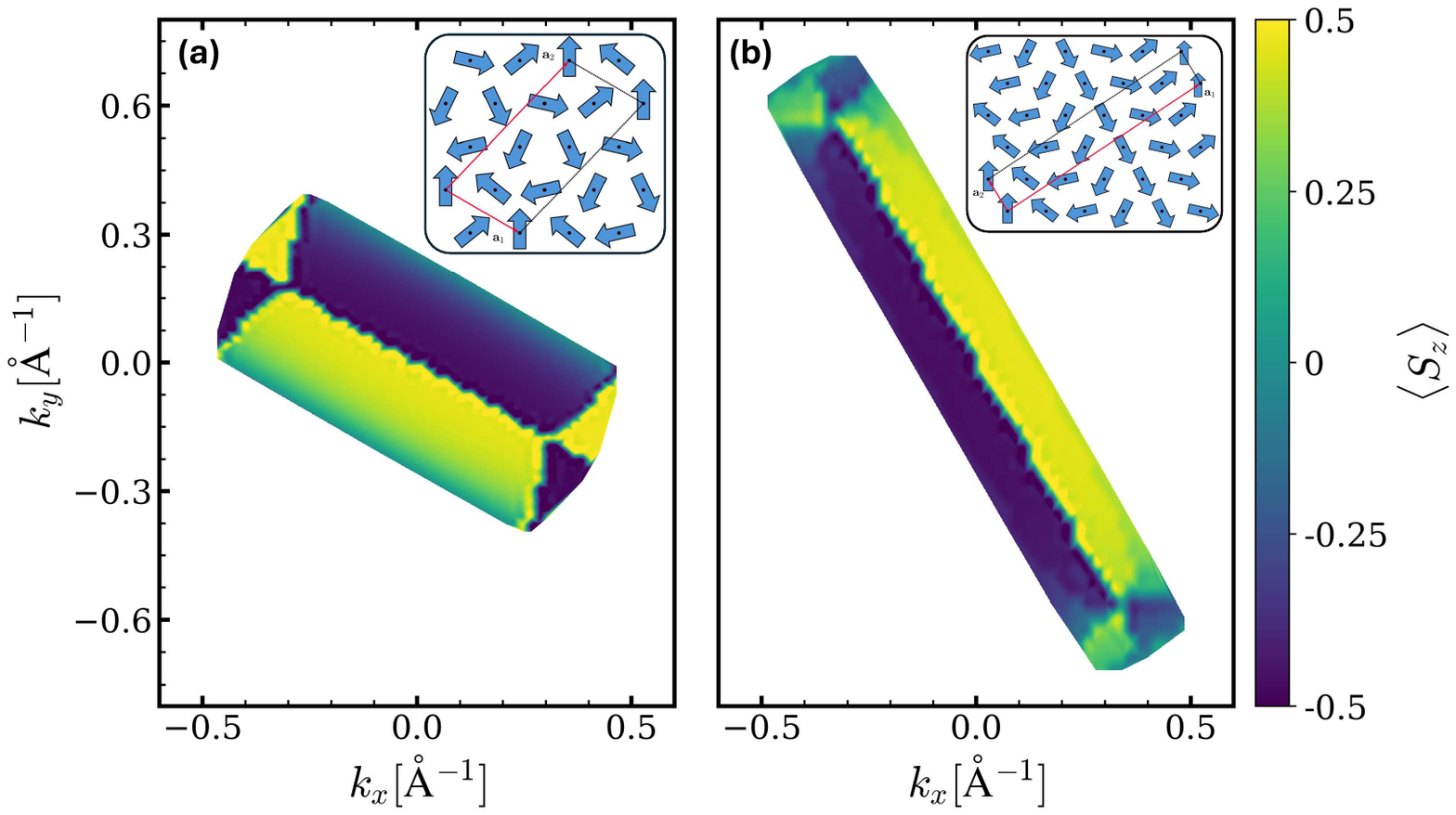}
\caption{\label{fig:NiI2_horizontal}Figure illustrating spin expectation value in the $\mathrm{NiI_2}$ super cell for (\textbf{a}) $\mathbf{Q}=(1/7, 1/7)$ and (\textbf{b}) $\mathbf{Q}=(1/7,0)$. The corresponding super cells are shown in the figure. }
\end{figure}

As an example of a more complex system, we consider the case of 2D $\mathrm{NiI_2}$, which have recently been shown to exhibit odd-parity magnetism and type II ferroelectricity driven by spiral order \cite{song_evidence_2022,song_electrical_2025}. While the bulk system has a ground state described by $\mathbf{Q}\sim(1/7, 0, 1.5)$ \cite{kuindersma_magnetic_1981}, the ordering vectors $\mathbf{Q}\sim(1/7, 0)$ and $\mathbf{Q}\sim(1/7, 1/7)$ are nearly degenerate in the monolayer \cite{sodequist_type_2023} and are difficult to distinguish experimentally in the monolayer limit where neutron diffraction is not applicable.  In addition, theoretical modeling is limited by the large super cells required to encompass the magnetic structure and the odd-parity bands were simulated using a $\mathbf{Q}\sim(1/3, 0)$ structure in Ref. \cite{song_electrical_2025}. In Fig. \ref{fig:NiI2_horizontal} we show the spin polarization of a valence band in NiI$_2$ for both the  $\mathbf{Q} =  (1/7, 1/7)$ and $\mathbf{Q} =  (1/7, 0)$ structures. Both cases yield a clear $p$-wave symmetry with nodal lines being orthogonal to $\mathbf{Q}$. The band polarization thus directly reflects the magnetic ordering vector, which may therefore be observable by spin-polarized ARPES measurements. We emphasize that the band structures and spin polarization in Fig. \ref{fig:NiI2_horizontal} have been calculated using the {\it primitive} unit cell within the GBT and then downfolded according to Eq. \eqref{Bloch-phase}.

In summary, we have presented a downfolding procedure that allows one to obtain the bands, spin polarization and wavefunctions of any helimagnet using only the primitive unit cell. We exemplified the approach for the simple case of MnI$_2$ and applied it to 2D NiI$_2$, which have a complicated magnetic unit cell with several near degenerate ordering vectors. It was shown that the spin polarization of the magnetic band structure can be used to extract the magnetic ordering vector and it therefore provides a direct route for determining the ordering vector from the band structure. Finally, the approach can be generalized to evaluate the matrix elements entering various response functions. For example, non-equilibrium spin densities induced by currents are often evaluated by means of Kubo-type expressions \cite{freimuth_spin-orbit_2014,li_intraband_2015} that involve Brillouin zone sums of momentum ($p_i$) and spin ($S_i$) matrix elements $\langle\tilde\psi_{\mathbf{K}m}|p_i/S_i|\tilde\psi_{\mathbf{K}n}\rangle$. It is straightforward to evaluate these based on calculations of $\psi_{\mathbf{q},\mathbf{k},m'}$ in the primitive unit cell: one simply identifies the set of $\{\mathbf{k}\}$ that downfolds to a given $\mathbf{K}$ and constructs the corresponding super cell wavefunctions from $\psi_{\mathbf{q},\mathbf{k},m'}$ using Eq. \eqref{Bloch_generalized}. In this respect, the GBT and the associated downfolding procedure may greatly accelerate the search for new helimagnetic materials for efficient non-relativistic charge to spin conversion.

\bibliography{apssamp}

\end{document}